# Hybrid Graphene Tunneling Photoconductor with Interface Engineering Towards Fast Photoresponse and High Responsivity

*Li Tao*, *Zefeng Chen*, *Xinming Li\**, *Keyou Yan,* and *Jian-Bin Xu\**

Department of Electronic Engineering, The Chinese University of Hong Kong, Shatin, N.T., Hong Kong SAR, China

*Email: xmli1015@gmail.com; jbxu@ee.cuhk.edu.hk

**Abstract:** Hybrid graphene photoconductor/phototransistor has achieved giant photoresponsivity, but its response speed dramatically degrades as the expense due to the long lifetime of trapped interfacial carriers. In this work, by intercalating a large-area atomically thin $MoS_2$ film into a hybrid graphene photoconductor, we have developed a prototype tunneling photoconductor, which exhibits a record-fast response (rising time ~17 ns) and a high responsivity (~$3\times10^4$ A/W at 635 nm and 16.8 nW illumination) across the broad spectral range. We demonstrate that the photo-excited carriers generated in silicon are transferred into graphene through a tunneling process rather than carrier drift. The atomically thin $MoS_2$ film not only serves as tunneling layer but also passivates surface states, which in combination delivers a superior response speed (~3 order of magnitude improved than a device without $MoS_2$ layer), while the responsivity remains high. This intriguing tunneling photoconductor integrates both fast response and high responsivity and thus has significant potential in practical applications of optoelectronic devices.

## 1. Introduction

Photodetectors with high speed and responsivity are of crucial importance for optoelectronic applications. Benefiting from their large surface area and unique electronic properties, two-dimensional (2D) materials are promising building blocks in optoelectronic devices.[1, 2] The gapless semi-metallic band structure of graphene gives rise to its high mobility, extremely broad optical absorption and ultrafast carrier dynamics,[3-6] making this emerging 2D material particularly attractive in photodetection applications.[7, 8] However, the responsivity of phototransistor comprised of pristine graphene is limited to the level of mA/W due to the low optical absorption rate and fast carrier recombination in graphene.[8, 9]

Recently researchers have proposed an effective method to improve the responsivity of graphene-based photodetector by adding a light absorption material, such as PbS quantum dots,[10] $Bi_2Te_3$,[11] perovskites,[12, 13] silicon (Si),[14, 15] carbon nanotubes,[16] atomically thin $MoS_2$,[17-18] and organic crystal,[19] as active layer and mainly remaining the graphene as the channel transport layer. This hybrid graphene photoconductor/phototransistor introduces a vertical built-in field at the interface for exciton dissociation. The photo-excited carriers in the active layer can be efficiently transferred into graphene, of which the electrical conductance is highly sensitive to external carrier injection.[20] Although this device configuration brings great enhancement of responsivity, the response speed of graphene-based photoconductor drastically degrades as the cost. The interface trap states, together with the drift time of photo-excited carriers in the depletion region, increase the response time into the range from several microseconds to several seconds.[10-19] This trade-off between response speed and responsivity makes it a great challenge to achieve fast and high-gain photoresponse simultaneously in graphene-based photodetectors, severely limiting their potential application for photodetection.

Interface passivation has been demonstrated to improve the performance of diode photodetector,[21-23] solar cell,[24-26] and tunneling filed-effect transistor (TFET),[27, 28] but has rarely been proved in a conductive mode photodetector, i.e., a photoconductor/phototransistor. Here we propose a high-performance atomically thin hybrid graphene tunneling photoconductor constructed with light absorption and interface passivation layers. This hybrid graphene photoconductor exhibits a response time of ~17 ns, which is improved by ~3 orders of magnitude compared with the device without $MoS_2$ intercalation and is at present the shortest one among those of hybrid graphene photoconductors/phototransistors, along with a high responsivity across a broadband spectral range (~$3\times10^4$ A/W at 635 nm illumination with 16.8 nW power). In our device, the Si serves as an optically active layer and $MoS_2$ film acts as a passivation layer for reducing surface states, suppressing carrier recombination at the interface, and offering an ultra-thin layer for the carrier tunneling. Carrier transfer process driven by ultra-fast quantum tunneling effect rather than by the carrier drift in the depletion region enables a superior response speed, while the responsivity remains high as a major merit of hybrid graphene photoconductor/phototransistor. The novel device designs and desirable results not only provide a unique structure for better understanding the photocarrier transfer and transport processes in 2D layered nanomaterials but also pave the way for the development of state-of-the-art graphene-based photodetector with both high speed and responsivity for practical applications.

## 2. Results and Discussion

### 2.1. Device Structure

**Figure 1a** schematically shows the structure of the hybrid graphene photoconductor, and **Figure 1b** presents the top view optical image of a typical device. A moderately n-doped silicon (n-Si, carrier density ~$10^{15}$ cm$^{-3}$) wafer was used as the starting substrate. A 100 nm $SiO_2$ thin layer was grown on the Si wafer and a 105 μm ×115 μm Si window was defined by $CF_4$ etching. Afterward, monolayer $MoS_2$ film grown by chemical vapor deposition (CVD) was transferred onto the Si groove region of the substrate, and a subsequent pattern process was carried out to enable the remaining $MoS_2$ cover the Si window. After monolayer graphene film was transferred onto the MoS₂/Si heterostructure, a pair of Au source/drain electrodes was thermally deposited atop graphene to form a channel with 125 μm in length. **Figure 1c** displays the Raman spectra taken in the vertical heterostructure area of the device (violet dashed rectangle in **Figure 1b**). Characteristic Raman signals of both MoS₂ and graphene were clearly detected. The feature peaks of MoS₂ at 384.6 cm$^{-1}$ ($E^1_{2g}$) and 403.1 cm$^{-1}$ ($A_{1g}$) with an 18.5 cm$^{-1}$ difference indicate the monolayer nature of the MoS₂ film.[29] The high quality of monolayer graphene was also confirmed by the absence of D band and the high intensity ratio of 2D and G peaks.[30] **Figure 1d** shows a typical optical image of large-scale continuous monolayer MoS₂ film before transfer. Photoluminescence (PL) measurement of the MoS₂ film (upper left inset of **Figure 1d**) reveals a direct optical bandgap of 1.82 eV, which reconfirms its monolayer feature.[31] High-resolution transmission electron microscopy (HRTEM) image in the upper right inset of **Figure 1d** presents a highly crystalline structure of monolayer MoS₂.

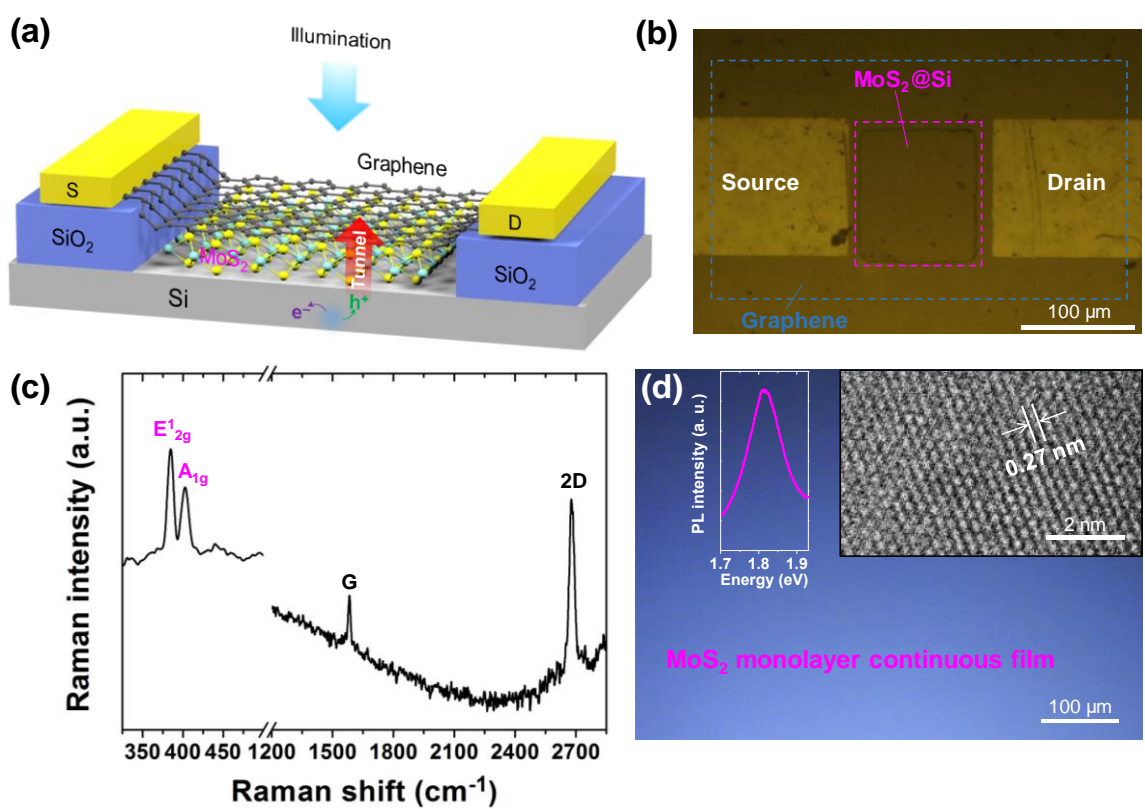

**Figure 1**. Schematic diagram and characterizations of the hybrid graphene photoconductor. (a) Schematic and (b) top view optical image of the device. (c) Raman spectra taken in the area within the violet dashed rectangle in (b). (d) Optical image of monolayer MoS₂ film before

transfer. Upper left inset: a typical PL spectrum of monolayer $MoS_2$ film with the peak at 1.82 eV. Upper right inset: HRTEM image of the $MoS_2$ film.

**2.2. Photoresponse Performance**

**Figure 2** provides the photoresponse characteristics of the hybrid graphene photoconductor. Typical illumination power-dependent response at a wavelength of 635 nm has been measured, as shown in **Figure 2a**. The device shows negative photoresponses because photo-excited holes are injected into n-type graphene while electrons are trapped in Si, lowering graphene's quasi-Fermi level through photo-gating effects.[10, 14, 32] Both $MoS_2$ and n-Si can bring n-dopants to graphene (see **Supporting Information S1**),[14, 18, 33] thus the Fermi level of graphene prior to being illuminated is above its Dirac point. The work function of graphene in the channel area is measured to be ~4.4 eV with a Kelvin probe force microscopy (KPFM) system (**Figure S1b**). The photocurrent $I_{ph}$, which is defined as the absolute difference between light current $I_{light}$ and dark current $I_{dark}$ ($I_{ph} = |I_{light} - I_{dark}|$), shows a monotonic increase as the light power $P$ increases. The responsivity, together with the photocurrent as functions of illumination power at $V_d = 5$ V are given in **Figure 2b**. The photoconductor shows a considerably high responsivity of ~3×10$^4$ A W$^{-1}$ at the light power of 16.8 nW, which far exceeds that in pure graphene device, and has the same order as that in graphene/Si photoconductor without $MoS_2$ intercalation (see **Supporting Information S2**). As the responsivity increases with decreasing light power, a higher responsivity is expected in ultra-weak illumination. The corresponding photoconductive gain $G$ is calculated to be ~5.8×10$^4$ by using the relation $G = (I_{ph}/q)/(P/h\nu)$, where $q$ is the elementary charge, $h$ is the Plank constant and $\nu$ is the frequency of incident light. Such an outstanding gain reveals that the photo-excited carriers can travel through the graphene channel for numerous times before recombination according to the definition of gain $G = \tau_{lifetime}/\tau_{tr}$, where $\tau_{lifetime}$ is the lifetime of injected carriers from Si and $\tau_{tr}$ is the carrier transit time across the channel.[34] The trapped photo-excited electrons in Si enable the excess hole conservation (with long $\tau_{lifetime}$) in graphene by capacitive coupling, leading to the carrier reinvestment in graphene channel and the ultrahigh gain in the photodetector.[10, 14, 35]

Wavelength-dependent photoresponses at 5 V bias were measured via quartz tungsten halogen lamp filtered with a monochromator. Normalized photocurrent versus light wavelength plotted in **Figure 2c** reveals a broad spectral response with considerable responsivity from visible light to the near-infrared region (~1120 nm). Relatively higher photoresponse than pure Si photodetector from violet to green is assigned to additional absorption by monolayer $MoS_2$.

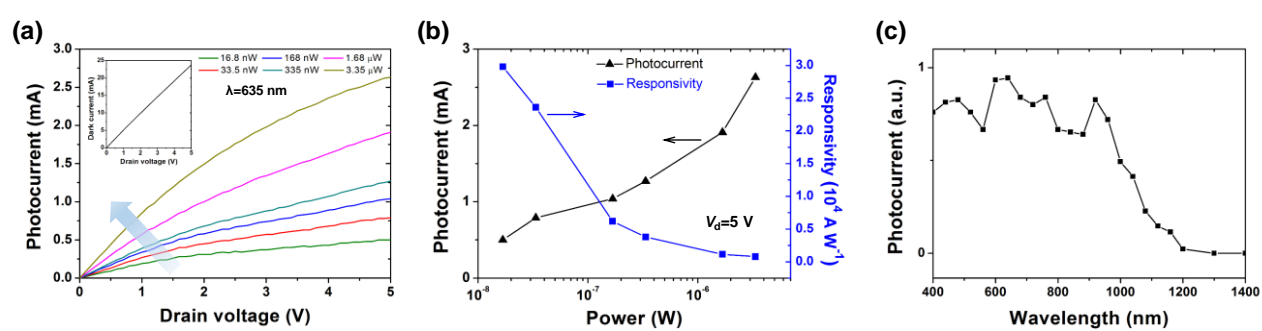

**Figure 2**. Photoresponse of the hybrid graphene photoconductor. (a) Photocurrent versus drain voltage under various light powers at 635 nm wavelength. The arrow indicates the direction along light power increase. The inset shows the dark current of the device. (b) Power-dependent photocurrent and photoresponsivity at 5 V drain voltage calculated from the data in (a). (c) Normalized photocurrent versus illumination wavelength.

### 2.3. Transient Performance

Temporal response of the hybrid graphene photoconductor was conducted under electrically modulated 635 nm light pulse having illumination power of 3.35 μW. **Figure 3a** shows the transient characteristics of the device at repeated frequency of 2 Hz, which demonstrates a highly repeatable and stable switching characteristics between light-off and light-on states. When the repeated frequency of light input increases to 100 kHz, the photoconductor can maintain its promising switching performance (**Figure 3b**). The fast rising edge followed by quick falling tail in the light-switching-on front results from a competition between hole drifting and electron diffusion processes in the depletion region.[15] The spikes in light switching points are due to the limitation of the pulse generator.

In order to further investigate the response speed of the device, which reveals the capacity to follow high-frequency optical signals, the detailed information of the rising and falling edges in a single switching cycle is provided in **Figures 3c** and **3d**, respectively. An ultra-fast response is found with the rising time (response time) $\tau_{\text{rise}}$ of ~17 ns and falling time (decay time) $\tau_{\text{fall}}$ of ~1.1 μs by fitting the rising and falling edges using exponential equations $I_{\text{ph}} = I_{\text{OFF}}(1 + A\exp\left(-\frac{t}{\tau_{\text{rise}}}\right))$ and $I_{\text{ph}} = I_{\text{ON}}(1 - B\exp\left(-\frac{t}{\tau_{\text{fall}}}\right))$, respectively, where $I_{\text{OFF}}$, $I_{\text{ON}}$, $A$ and $B$ are fitting parameters. In contrast to the graphene/Si photoconductor without MoS$_2$ layer intercalation which gives a response time of 12 μs (inset in **Figure 3c**), this hybrid graphene photoconductor with MoS$_2$ interface layer greatly enhances the response speed by ~3 orders of magnitude. Transient response measurements under infrared light (905 nm) were also performed (see **Supporting Information S3**), from which similar fast response ($\tau_{\text{rise}}$~35 ns) can be identified. Note that some oscillation signals with a period of ~28 ns occur at the rising and falling edges; it is suggested to be the intrinsic RC oscillation of the operating circuit. By measuring the resistance $R$ and capacitance $C$ of the operating circuit (graphene channel) (2.3×10$^2$ Ω and 2.1×10$^{-11}$ F, respectively), we obtain the calculated RC time constant $T_{\text{RC}} = 2\pi RC \approx 30$ ns, which is self-consistent with the directly measured value from the oscilloscope. The phenomenal response speed of the tunneling graphene-based photoconductor far exceeds those in previous reports on hybrid graphene photoconductors/phototransistors, and is even close to the limitation of intrinsic circuit oscillation. Furthermore, in stark contrast to commercial Si photoconductors/phototransistors of which response times are around several tens of microseconds (e.g., OP602TX Si phototransistor from OPTECK with $\tau_{\text{rise}}$~20 μs),[36] this hybrid graphene photoconductor overcomes the limitation of pure Si device by enhancing the response speed of about 3 orders. This results from the photo-excited carriers in our device transiting across the graphene channel, which is a much faster process than the carriers transiting across the Si channel in pure Si photoconductors. Such fast photoresponse makes this device highly promising in high-frequency optical sensing applications such as military warning, high-speed imaging and chemical sensing.

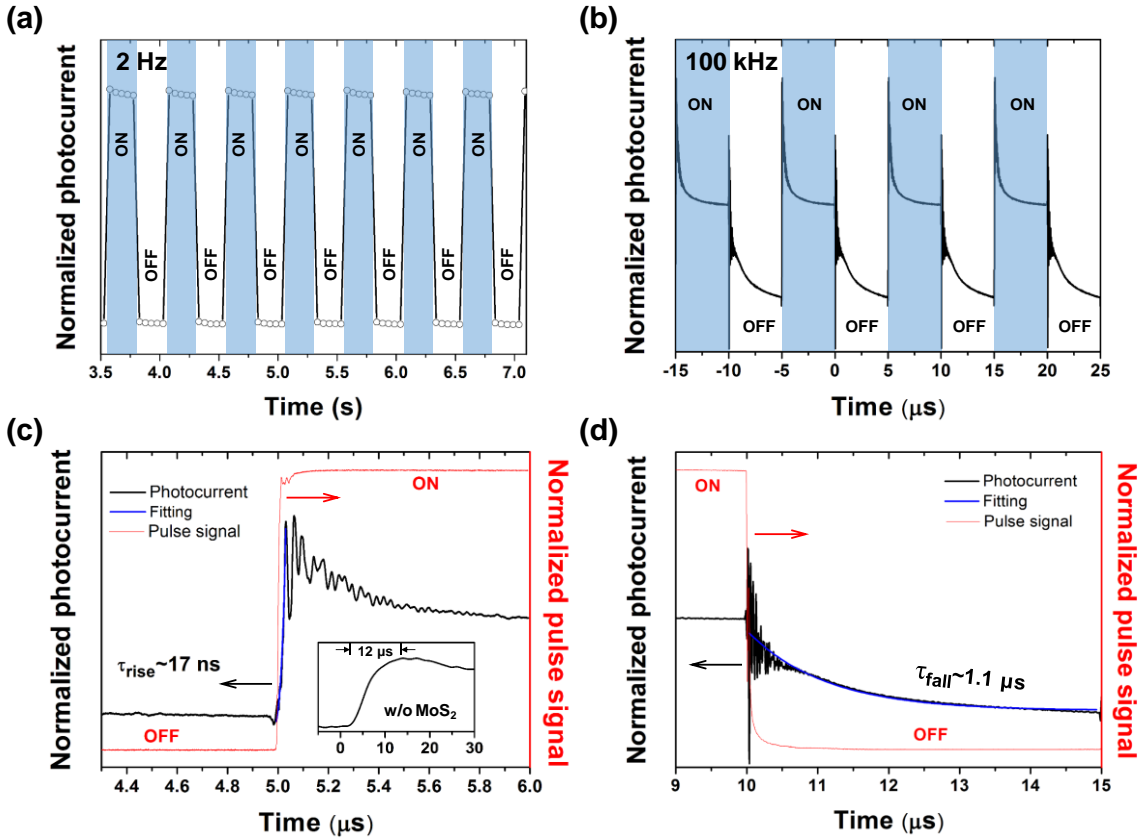

**Figure 3**. The temporal performance of the hybrid graphene photoconductor. Switching characteristics of the device under (a) 2 Hz and (b) 100 kHz square-wave light pulse (635 nm). Enlarged views of (b) at light being switched on (c) and light being switched off (d). Inset in (c): rising edge of the device without $MoS_2$ layer intercalation.

## 2.4. Photocarrier Tunneling and Interface Passivation

*2.4.1. Photocarrier Transport Process*

We have proposed a synergistic mechanism to explain the extraordinary performance of the hybrid graphene photoconductor with $MoS_2$ interface layer, which will be demonstrated in the following. **Figure 4a** illustrates the band diagram of the device. A Schottky barrier is formed at the interface between graphene and Si. Electron and hole pairs (EHPs) are generated in the depletion region under illumination. The excess holes in n-Si result in a small hole quasi-Fermi level difference between graphene and Si ($\Delta E_{\text{Fp}}$), acting as the driving force for hole transport.[25, 37] The built-in field separates the EHPs and consequently the holes are transferred into graphene, leading to the conductance reduction of graphene. It is estimated that the quasi Fermi level of graphene is pulled down by 9.2 meV due to the photo-excited holes injection under 3.35 μW illumination and the built-in barrier height $q\phi_{bi}$ prior to being illuminated is ~0.15 eV (see **Supporting Information S4**). This Fermi level lowing caused by illumination is also detected by the Raman G peak redshift in n-type graphene with increasing laser power (**Figure S4b**). Here the MoS$_2$ film plays several crucial roles in device operating. First, MoS$_2$ passivates the interface of the Schottky junction to reduce trap states and to further suppress carrier recombination. Meanwhile the ultra-flat MoS$_2$ underneath graphene allows the intrinsic electronic properties to be retained in graphene.[38] Second, the relatively wide electronic bandgap of monolayer MoS$_2$ (~2.4 eV) results in a straddling gap at MoS$_2$/Si interface, namely, type-I heterostructure.[39, 40] In this band alignment, photo-excited holes are accumulated at the interface triangular well (green circles in **Figure 4a**), while electrons are blocked from transiting towards graphene. The valence band offset at MoS$_2$/Si ($\Delta E_V$) around 1.1 eV[39] provides a barrier to prohibit the thermionic emission for the photo-excited holes (thermal energy at 300 K is ~26 meV), thus suppressing the thermal noise. Third, the atomic thickness of MoS$_2$ enables the carrier wave function in Si to interact with that in graphene, causing the carrier quantum tunneling through MoS$_2$ layer. The band bending at MoS$_2$/Si interface (**Figure 4a**) generates a strong built-in field that further enhances the hole tunneling.

*2.4.2. Photocarrier Tunneling*

The photo-excited carriers (holes) in the device are injected into graphene via an ultra-fast tunneling process. The response time $\tau_{rise}$ is affected by the drift time for photo-excited carriers to travel from depletion region to the Schottky junction interface ($\tau_{drift}$), the average time for carriers to transit through interface trap states ($\tau_{trap}$), the average time for carriers to tunnel through MoS$_2$ film ($\tau_{tunnel}$) and carrier transit time in graphene channel ($\tau_{tr}$). $\tau_{tr}$ is estimated to be $\tau_{tr} = \frac{L}{\mu_{graphene}V_d/L} \approx 10$ ns, where $L$ is the channel length (125 μm) and $\mu_{graphene}$ is carrier mobility in graphene which has been derived to be ~ $3.1\times10^3$ cm$^2$V$^{-1}$s$^{-1}$ (see **Supporting Information S4**). $\tau_{trap}$ is greatly lessen with MoS$_2$ passivation, as discussed above. $\tau_{drift}$ is also reduced when MoS$_2$ layer is used since MoS$_2$ shares the depletion region of the Schottky junction and makes the length of remaining depletion region in Si shorter. $\tau_{tunnel}$ is highly sensitive to the tunneling barrier (MoS$_2$ film) thickness $d$, which is written as below based on the Wentzel-Kramers-Brillouin approximation:[34, 41]

$$\tau_{tunnel} \propto 1/T_{tunnel} \propto \exp\left[2d\sqrt{\left(\frac{8\pi^2 m_p^*}{h^2}\right)q\phi_{tunnel}}\right] \qquad (1)$$

where $T_{tunnel}$ is the carrier tunneling probability, $m_p^*$ is the effective mass of the accumulated hole in Si, and $q\phi_{tunnel}$ is the average barrier height for tunneling. The carrier tunneling phenomenon originates from the wave nature of carriers in quantum mechanics, and the tunneling time $\tau_{tunnel}$ is not determined by the classic transit time process (i.e., $\tau = d/v$, where $v$ is the carrier velocity).[34] Tunneling time can be very short when the tunneling barrier is ultrathin (<2 nm), typically in the timescale of subnanosecond.[41] The high speed response in the photoconductor benefits dominantly from this ultra-fast tunneling process when the MoS$_2$ layer is thin enough. Additionally, graphene based Schottky junction enables the tunneling effect to occur in relatively small built-in barrier while conventional TFETs require a gate voltage for band-to-band tunneling.[28]

To certify the tunneling mechanism described above, we investigate the thickness effect of the MoS$_2$ passivation layer on the photoconductor performance. **Figure 4b** depicts the energy-band diagram of the hybrid graphene photoconductor with multilayer MoS$_2$ interface layer. As the tunneling barrier gets thicker, the carrier tunneling probability $T_{tunnel}$ drastically reduces (Eq. (1)). Correspondingly, not only the tunneling time $\tau_{tunnel}$ which is inversely proportional to $T_{tunnel}$ becomes much longer, but also the total population of photo-excited carriers which can tunnel into graphene decreases. We have fabricated additional devices with multilayer MoS$_2$ film (~4 layers, confirmed by Raman spectrum, **Figure S5a**), with all the other factors unchanged. Temporal response measurements show that the photoconductor with thicker interface layer presents a response time of ~500 ns (**Figure 4c**), revealing that larger tunneling barrier thickness $d$ causes a longer $\tau_{\text{tunnel}}$ (Eq. (1)). Moreover, the photoresponsivity in the photoconductor with multilayer MoS$_2$ is much smaller than that of device with monolayer MoS$_2$ (**Figure 4d**), which well agrees with the above analysis on the injected carrier population decrease (note that the increased light absorption by MoS$_2$ is limited).[42]

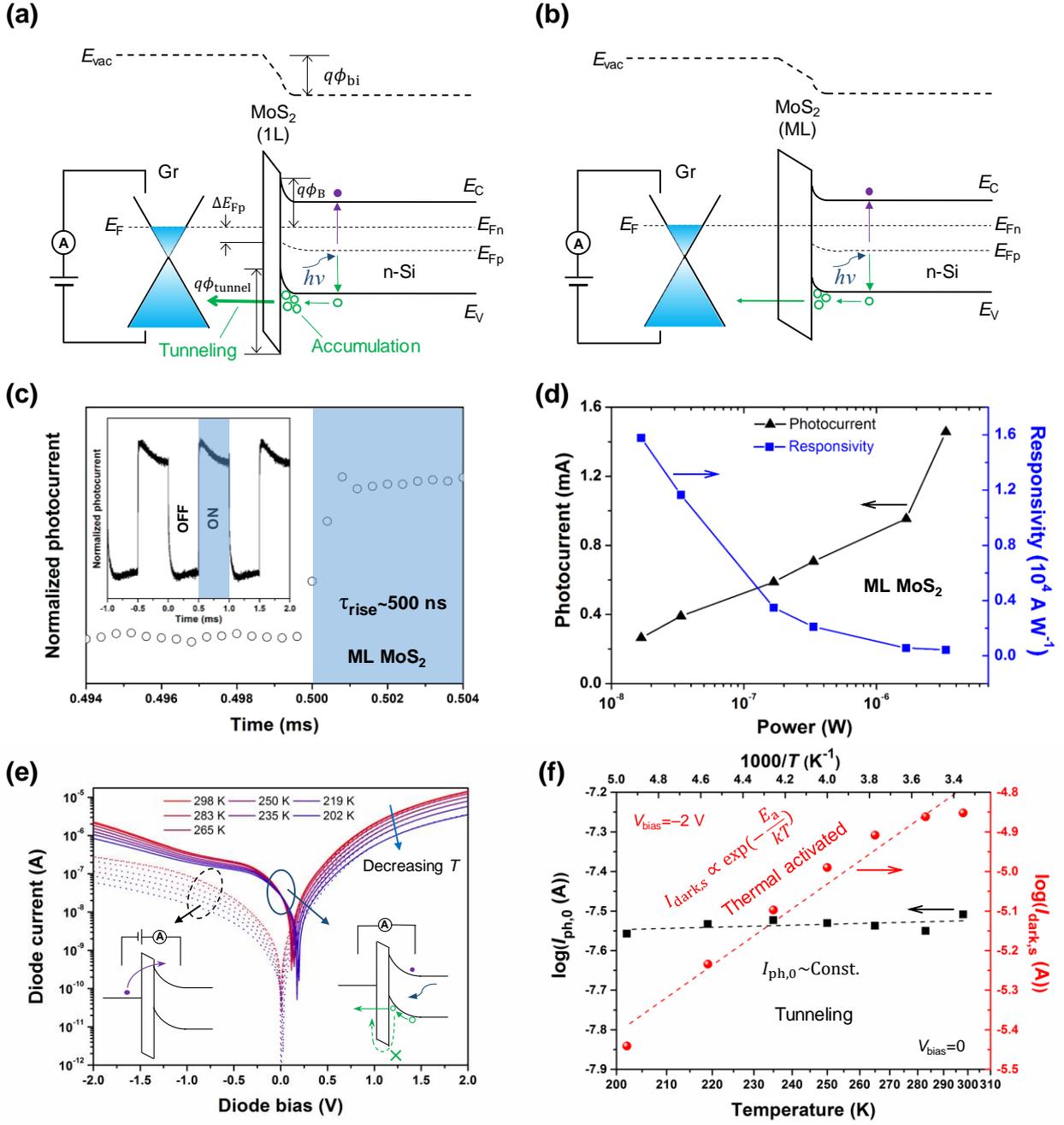

**Figure 4.** Photoexcited carrier tunneling process in the hybrid graphene photoconductor. Energy band diagram of the devices formed with monolayer (1L) (a) and multilayer (ML) (b) MoS$_2$ films. The number of the green circles indicates the relative population of accumulated carriers. The accumulated carriers can tunnel to graphene through MoS$_2$ layer, as shown by the green lines. The sizes of the green lines indicate the relative tunneling probability. A thicker tunneling barrier results in a smaller tunneling probability. (c) Transient characteristics of the hybrid graphene photoconductor with ML MoS$_2$ under 635 nm illumination, showing a rising time of ~500 ns. Inset is the switching performance over 3 periods of square-wave modulation. (d) Photocurrent and responsivity as functions of the illumination power of the device with ML

MoS$_2$. (e) Current versus bias curves of the device operating in diode mode in the dark (dashed curves) and under 635 nm (3.35 µW) illumination (solid curves) at various temperatures. Insets schematically show the carrier transport in dark (reverse bias) and under illumination (zero bias), respectively. (f) Zero-bias photocurrent ($I_{\text{ph},0}$) and reverse saturation dark current at –2 V ($I_{\text{dark,s}}$) as functions of temperature (Arrhenius plot) extracted from (e). The dashed lines show the linear fittings.

To further understand the photocarrier tunneling process, we investigate the temperature-dependent carrier transport in the device operating in diode mode (i.e., the source and drain electrodes are on graphene and Si, respectively). **Figure 4e** presents the photocurrent and dark current at various temperatures. Under the dark condition, the carriers have to surmount the Schottky barrier ($q\phi_{\text{B}}$) for transport (left inset in **Figure 4e**), and the dark current obeys to the traditional thermionic emission theory for Schottky diode:[34]

$$I_{\text{dark}} = AA^*T^2 \exp\left(-\frac{q\phi_{\text{B}}}{kT}\right)\left[\exp\left(\frac{qV}{\eta kT}\right) - 1\right] \quad (2)$$

where A is the junction area, A$^*$ is the effective Richardson constant, $T$ is the absolute temperature, $k$ is the Boltzmann constant, and $\eta$ is the ideality factor. The reverse saturation current at diode bias $V_{\text{bias}} = -2$ V ($I_{\text{dark,s}}$) as a function of temperature is plotted in red circles in **Figure 4f** (Arrhenius plot). A linear $\log(I_{\text{dark,s}})$-$1/T$ dependence is found, demonstrating the carrier transport in the dark is a thermally activated process with the express $I_{\text{dark}} \propto \exp(-\frac{E_{\text{a}}}{kT})$, where $E_{\text{a}}$ is the activation energy.[34] When illuminating the diode at zero bias, the current is fully contributed by photo-excited carriers ($I_{\text{ph},0} = I_{\text{light},0}$ at $V_{\text{bias}} = 0$), and the holes tunnel through MoS$_2$ layer into the graphene rather than being thermally activated (right inset in **Figure 4e**). As illustrated in **Figure 4f**, $I_{\text{ph},0}$ at zero bias is temperature insensitive. The short-circuit hole tunnel photocurrent is given by:[37]

$$I_{\text{ph},0} = \frac{4\pi m_{\text{p}}^* q(kT)^2}{h^3 N_{\text{V}}} p_{\text{s}} \exp(-\sqrt{q\phi_{\text{tunnel}}}d)\left[1 - \exp\left(-\frac{\Delta E_{\text{Fp}}}{kT}\right)\right] \quad (3)$$

where $N_V$ is the effective density of states in the valence band of Si, and $p_s$ is the concentration of holes at the interface. This expression indicates a non-thermally-activated process of the photo-excited carrier since $\Delta E_{Fp}$ promotes the carrier tunneling rather than setting up an energy barrier. From the Richardson plot of $I_{ph,0}$ ($\log(I_{ph,0}/T^2)$-$1/T$ plot) as illustrated in **Figure S6**, one can identify that $I_{ph,0}/T^2$ increases with decreasing temperature, which is in good coincidence with Eq. (3). The above analysis of photo-excited carrier transport at zero bias in the graphene/Si diode with MoS$_2$ layer can be fully applied to the device operating in photoconductor mode since there is no external vertical bias across the photoconductor (**Figure 1a**).

*2.4.3. Interface Passivation*

In previous studies of graphene photoconductor/phototransistor hybridized with a semiconducting active layer, the photo-excited carriers were transferred to graphene driven by the built-in field. During this process, the carriers can be trapped by interface states, thus resulting in long carrier transfer time. Here, by inserting an atomically thin MoS$_2$ layer into hybrid graphene photoconductor, the interface traps are passivated and thus the photo-excited carriers can be efficiently transferred into graphene through the ultra-fast tunneling process. We further investigate the MoS$_2$ passivation effect by the noise spectra of the devices. It is widely accepted that the noise in metal-oxide-semiconductor transistors can be explained by the trap charge fluctuation mechanism.[43] The interface trap states are considered to be one of the dominant contributions to the low-frequency 1/$f$ noise in monolayer graphene devices.[44-46] It is observed that the graphene-based photoconductor with MoS$_2$ interface layer performs more than one order of magnitude lower noise power density ($S_I/I^2$) when compared with the device without MoS$_2$ layer (**Figure 5a**). The average noise amplitude $A_{noise}$, which is defined as $A_{noise} = \frac{1}{N}\sum_{k=1}^{N} f_k (S_I/I^2)_k$, can quantitatively probe the level of 1/$f$ noise for a given frequency range.[44] For devices without and with MoS$_2$ passivation, $A_{noise}$ is calculated to be 1.5×10$^{-8}$ and $2.4×10^{-10}$, respectively at low frequencies (4 - 191 Hz). This large noise reduction indicates that the interface traps are greatly removed by $MoS_2$ passivation, demonstrating that $MoS_2$ serves as perfect substrate for graphene.[38] By designing the novel graphene-based photoconductor with interface engineering, we have approached the highest response speed among hybrid graphene photoconductors/phototransistors. For detailed comparison, we have summarized the figures of merit (response time and responsivity) for some previously reported hybrid graphene photoconductors/phototransistors and our device, as shown in **Figure 5b**. [10-19]

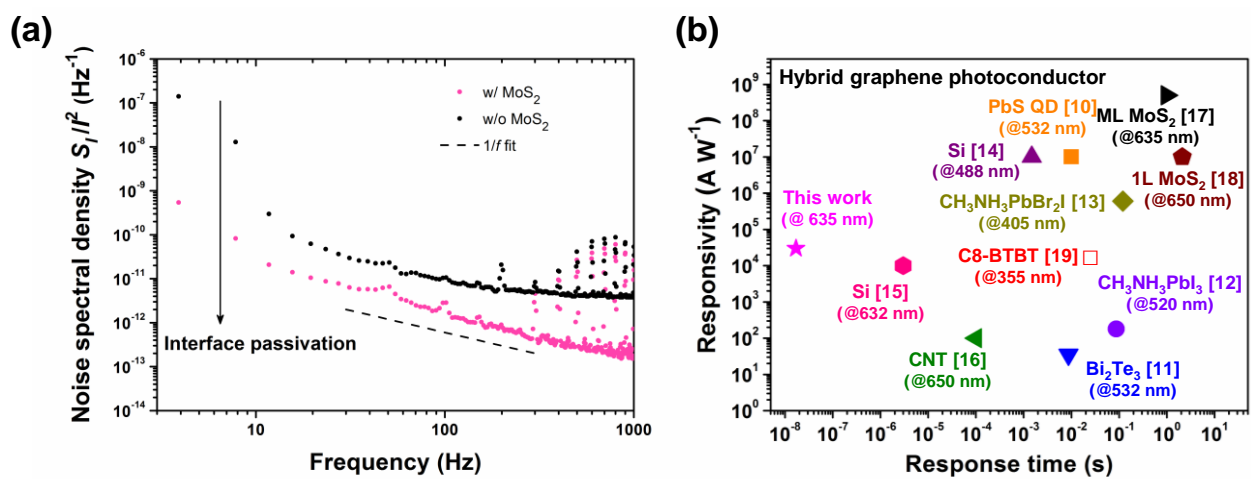

**Figure 5.** (a) Normalized noise spectral density ($S_I/I^2$) as a function of frequency (*f*) for the hybrid graphene photoconductor with and without monolayer $MoS_2$ passivation. Dashed line gives the guide to eyes for 1/*f* noise. Sharp peaks at 50 Hz and its harmonics are induced by power source. (b) Comparison between response time (rising time) and responsivity of our device and those of typical reported hybrid graphene photoconductors/phototransistors, with the photoactive material adopted in the device, the corresponding reference number (in square brackets) and the operation wavelength (in parentheses) in each labeled text. Abbreviations: quantum dot (QD); carbon nanotube (CNT); dioctylbenzothien-obenzothiophene (C8-BTBT).

### 2.5. Flexible Device

Furthermore, this hybrid graphene photoconductor does not require electrical gating to operate, which makes it possible to obtain flexible device if Si thin film (thickness ~200 nm) exfoliated from silicon-on-insulator is adopted (see **Supporting Information S7**). Here we fabricated a flexible device on a polyimide substrate, as manifested in **Figure S7a**. This flexible device shows the high reliability of photoresponse under 635 nm illumination after mechanical bending for 100 times (**Figures S7b-S7c**). The responsivity of the flexible photoconductor under 168 nW light power is ~$1\times10^3$ A/W, and the response time $\tau_{rise}$ is around 0.2 s. The flexible device exhibits competitive performance when comparing to previously reported flexible photodetector based on graphene[47] and other 2D materials.[48] Due to the large amount of traps in polyimide and Si thin film, the flexible device cannot response as fast as the rigid one does at present.

3. Conclusion

In summary, we developed a novel interface engineered hybrid graphene photoconductor with a record-fast response within 17 ns and a high photoresponsivity across a wide spectral band. The atomically thin $MoS_2$ film in the device not only passivates interface traps but also acts as a barrier for photo-excited carrier tunneling into graphene, which accounts for the remarkable response speed. We have investigated the tunneling mechanism in detail and identified the importance of incorporating tunneling barrier in atomic thickness. Temperature-insensitive short-circuit photocurrent in graphene/Si diode with $MoS_2$ interface layer provides strong evidence for the photo-excited carrier tunneling in the device. Also, the interface passivation by $MoS_2$ is demonstrated through noise analysis. This demonstrated graphene-based photoconductor with interface engineering opens a pathway towards ultra-fast, high responsivity and broadband photodetection with the large photo-active area, therefore providing a myriad of opportunities for future study and practical applications of 2D materials.

4. Methods

*CVD growth of $MoS_2$*: Monolayer/multilayer $MoS_2$ films were fabricated using a home-built CVD system.[49] A quartz boat holding ammonium molybdate powder (6 mg) for producing uniform $MoO_3$ during growing was put in the center of a 2-inch CVD quartz tube. $SiO_2$/Si substrates treated with acetone and $O_2$ plasma were placed next to the boat. Another quartz boat holding precursor of S powder of 0.7 g was placed 30 cm away from ammonium molybdate. The reaction furnace was heated to 800 °C at the rate of 15 °C $min^{-1}$ and S powder was heated to 180 °C when the reaction zone reached the temperature of 650 °C with a 50 sccm Ar flow at the pressure of ~100 Pa. After ~30 min $MoS_2$ growing, the furnace was cooled down to room temperature and $MoS_2$ films with high quality were deposited on the substrates.

*Characterization*: A Renishaw Spectroscope with 514.5 nm laser excitation was applied to perform Raman and PL measurements. The size of laser spot was ~1 μm. HRTEM measurements were conducted via FEI Technai F20 system. KPFM measurements were performed using a Bruker Dimension Icon system.

*Photoresponse measurements*: A Keithley 4200 Semiconductor Characterization System was applied in the measurements. Lasers with 635 nm and 905 nm wavelengths and a series of attenuation slices were used to generate constant illumination with various powers. A quartz tungsten halogen lamp housing (Newport 66882) combined with a monochromator was used to generate illuminations with specific wavelengths (400 nm-1500 nm). The filtered light power is estimated to be ~2 nW at 635 nm wavelength. To make sure the lasers only illuminate the active layer area (channel) of the device, a black mask was applied to block the other region of the device. A square-wave pulse generator was used to modulate the laser to give switching illumination. An oscilloscope was connected in series with the device to investigate the switching characteristic at high-frequency optical signals. Noise analysis was conducted via a Stanford Research Systems SR760 FFT Spectrum Analyzer.

**Acknowledgements**
The authors are thankful to Dr. Hui Yu for valuable advice and Mr. Guodong Zhou and Mr. Ningqi Luo for technical support. The work is in part supported by Research Grants Council of Hong Kong, particularly, via Grant Nos. N_CUHK405/12, AoE/P-02/12, 14207515, 14204616, and CUHK Group Research Scheme. J. B. Xu and X.M. Li would like to thank the National

Natural Science Foundation of China for the support, particularly, via Grant Nos. 61229401 and 51402060. X.M. Li also thanks the Postdoctoral Fellowship by CUHK.

**Contributions**

L.T. designed the device structure. X.M.L. and J.B.X. supervised the project. L.T. did the material synthesis and characterization. L.T. and Z.F.C. fabricated the device and performed the photoresponse measurements. L.T., Z.F.C., X.M.L., K.Y.Y. and J.B.X analyzed the data. L.T. wrote the manuscript. L.T., X.M.L., K.Y.Y. and J.B.X. revised the manuscript. All the authors discussed the results and commented on the manuscript.

Supporting Information

# Hybrid Graphene Tunneling Photoconductor with Interface Engineering Towards Fast Photoresponse and High Responsivity

*Li Tao*, *Zefeng Chen*, *Xinming Li\**, *Keyou Yan,* and *Jian-Bin Xu\**

Department of Electronic Engineering, The Chinese University of Hong Kong, Shatin, N.T., Hong Kong SAR, China

*Email: xmli1015@gmail.com; jbxu@ee.cuhk.edu.hk

## S1. Substrate-induced Fermi level difference of graphene

**Figure S1a** shows the transfer curves of graphene on SiO$_2$/Si substrate and that on MoS$_2$/SiO$_2$/Si. Pristine CVD graphene on SiO$_2$ presents a p-type behavior[S1] (the conductive neutral point at 90 V) while MoS$_2$ underneath introduces n-dopants to graphene[S2] (conductive neutral point shifted to –12 V). n-Si can also bring n-dopants to graphene, which has been well discussed by Liu et al.[S3] The surface potential differences (SPDs) can be directly detected from the KPFM cross-sectional profiles (**Figure S1b**). The SPD between the channel graphene on the MoS$_2$/Si heterostructure area (G@MoS$_2$@Si) and the graphene on a gold electrode (G@Au) is around 309 meV while the SPD between G@Au and pure Au is about 87 meV. Therefore, by considering the work function of gold $W_{Au} = 4.8$ eV, we can obtain the work function of the channel graphene $W_{graphene} = 4.8 \text{ eV} - (0.309 \text{ eV} + 0.087 \text{ eV}) = 4.4 \text{ eV}$.

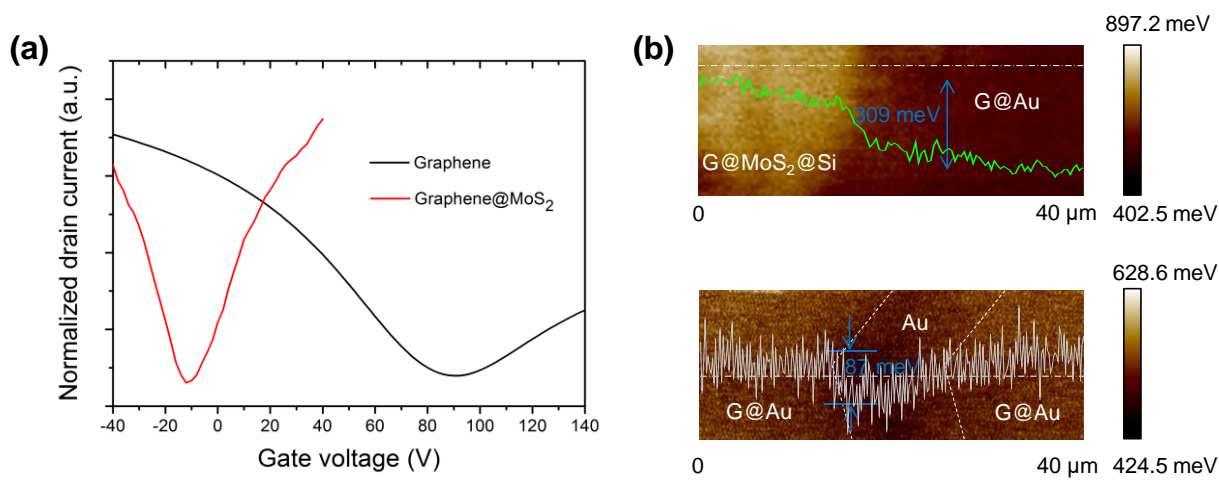

**Figure S1.** (a) Transfer curves of graphene on SiO$_2$/Si and graphene on MoS$_2$/SiO$_2$/Si at a drain voltage 1V. (b) The surface potential of the channel graphene in the device measured by KPFM. The green line indicates the surface potential difference between the channel graphene on the MoS$_2$/Si heterostructure area (G@MoS$_2$@Si) and the graphene on a gold electrode (G@Au). The gray line indicates the surface potential difference between the graphene on gold electrode and the pure gold. The horizontal white dashes are the positions for the KPFM cross-sectional profiles.

## S2. Power-dependent photoresponse of graphene/Si photoconductor without MoS$_2$ intercalation

Controlling devices of graphene/Si photoconductor without MoS$_2$ intercalation were fabricated (other factors remained unchanged). Photocurrent together with photoresponsivity as shown in **Figure S2** illustrate slightly higher values than those of the device with MoS$_2$ layer discussed in the main manuscript. The small photocurrent decreasing is attributed to 1) the less absorption efficiency of Si induced by MoS$_2$ blocking; and 2) the non-unity tunneling probability (<100%) of photo-excited carriers in Si to tunnel into graphene. Nevertheless, this responsivity degrading caused by MoS$_2$ intercalation is limited (from $4.7 \times 10^4$ A/W to $3 \times 10^4$ A/W), while the response speed improvement induced by MoS$_2$ is phenomenal (about 3 orders of magnitude), as discussed in the main manuscript.

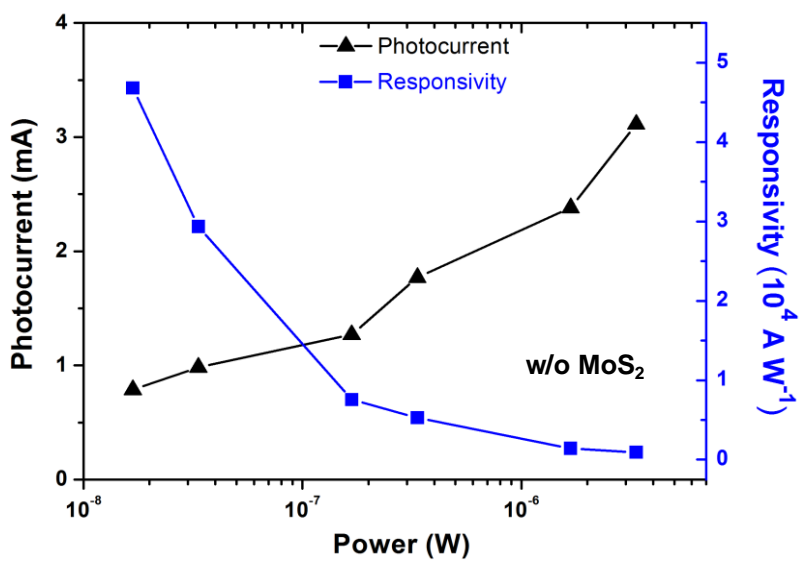

**Figure S2.** Photocurrent and photoresponsivity as functions of illumination power at 635 nm of graphene/Si photoconductor without MoS$_2$ intercalation.

## S3. Switching characteristics of the device under infrared illumination

Transient response measurements under infrared light (905 nm, 6.7 µW) were taken using 1 kHz pulse signals. The results shown in **Figure S3** reveal that the photoconductor performs ultra-fast response within 35 ns under infrared illumination, which is similar to the case of visible light illumination.

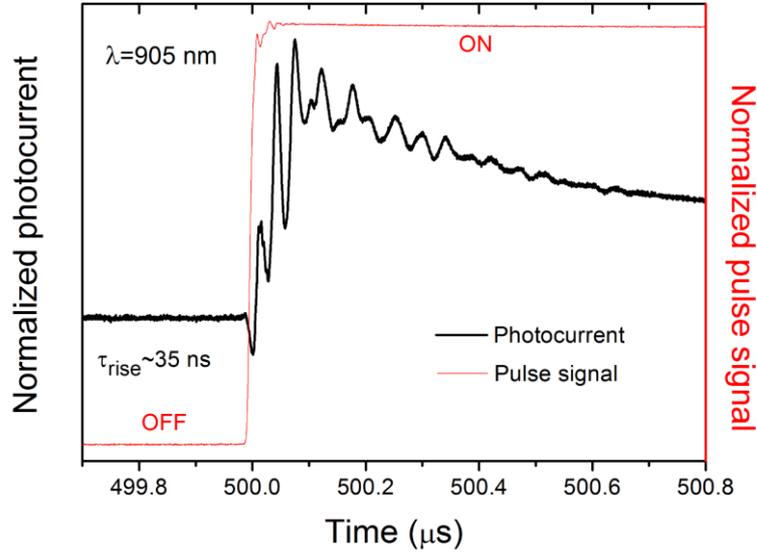

**Figure S3.** Switching characteristics of the hybrid graphene photoconductor under 905 nm illumination.

**S4. Fermi level shifting of graphene induced by photo-excited carrier injection and the built-in barrier height**

Because of the low density of states near the Dirac point in graphene, its quasi-Fermi level $E_F'$ is easily effected by excess carrier injection according to the equation[S4]

$$E_F' = \pm \hbar v_F k_F' = \pm \hbar v_F \sqrt{\pi(n_0 + \Delta n)} \tag{S1}$$

where $v_F \approx 10^6$ m/s is the Fermi velocity of the massless Dirac fermion in graphene, $k_F'$ is the Fermi wavevector magnitude, $n_0$ is the carrier density in graphene at equilibrium and $\Delta n$ is the excess carrier density due to illumination. $E_F'$ takes reference to the Dirac point level of graphene, and will be a positive/negative value if the graphene is conducted by electron/hole. The resistance of graphene channel under illumination $R_\text{channel}$ can be expressed as

$$R_\text{channel} = \frac{L/W}{(n_0+\Delta n)q\mu_\text{graphene}} \tag{S2}$$

where $L$ and $W$ are the length and width of the conductive graphene channel, respevtively, and $\mu_\text{graphene}$ is the mobility of carrier in graphene. In our device, $L/W$ is about 0.2. $R_\text{channel}$ under various illumination power is calculated from *I-V* curves in **Figure 2a**. From Eq. (S2), we can estimate the total conductive electron density in graphene $n = n_0 + \Delta n$ at different illumination powers. From Eqs. (S1-S2), the illumination power-dependent $E'_\text{F}$ can be calculated from the relation

$$E'_\text{F} = \pm\hbar v_\text{F}\sqrt{\pi\frac{L/W}{q\mu_\text{graphene}R_\text{channel}}} \tag{S3}$$

By considering that the intrinsic work function (Dirac point level) of graphene $W_\text{D} = 4.56$ eV, we can get the Fermi level of graphene in the photodetector prior to being illuminated $E_\text{F} = W_\text{D} - W_\text{graphene} = 0.16$ eV. As the $R_\text{channel}$ under no illimination $R_\text{channel,0} = 210$ Ω, we can get the mobility of graphene from Eq. (S3): $\mu_\text{graphene} \approx 3.1\times10^3$ cm$^2$V$^{-1}$s$^{-1}$. Therefore, all the constants in Eq. (S3) are determined and the equation can be written as $E'_\text{F} = \frac{2346 \text{ meV}}{\sqrt{R_\text{channel}/\Omega}}$. The illumination power-dependent carrier density and quasi Fermi level of graphene are plotted in **Figure S4a**, showing that as the light power increases to 3.35 μW, $E'_\text{F}$ is pulled down from 161.8 meV to 152.6 meV. Raman G peak shift is known to probe the doping level in graphene.[S5] We have taken Raman measurements of graphene G band in the device with various Raman laser powers, as shown in **Figure S4b**. As the laser power increases from 10% to 100%, the G peak shows a redshift from 1586.2 cm$^{-1}$ to 1582.3 cm$^{-1}$, indicating the graphene is getting less doped.

The surface potential of n-Si adopted in the device is around 4.27 eV, therefore the initial built-in barrier height $q\phi_\text{bi}$ of the device can be estimated to be 0.13 eV. Capacitance-voltage (*C-V*) measurement of the barrier without illumination, as is shown in **Figure S4c**, indicates the barrier height is ~0.15 eV, which is consistent with the estimation from KPFM measurements.

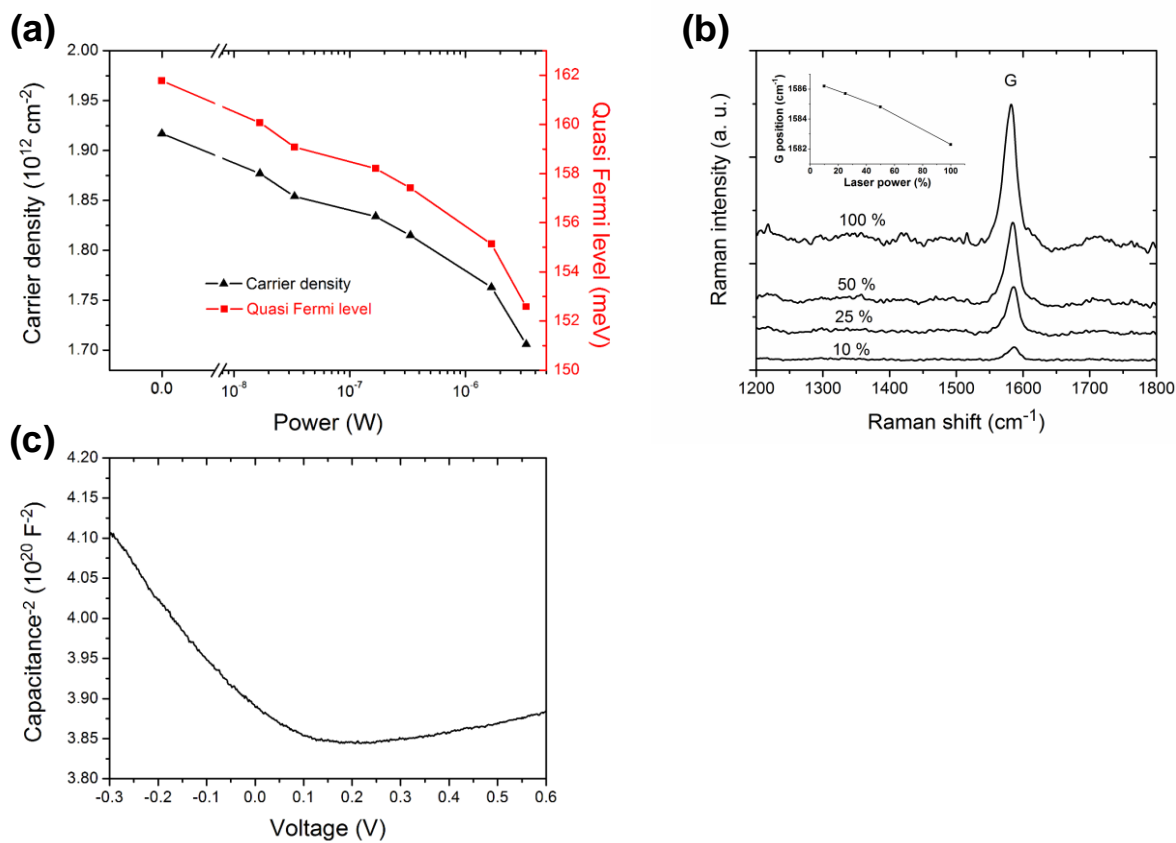

**Figure S4.** (a) Illumination power-dependent carrier density and quasi-Fermi level of graphene in the device. (b) Raman G band spectra of graphene in the device with various Raman laser powers. Inset is the G peak position as a function of laser power. (c) Capacitance-voltage characteristic of the graphene/Si Schottky junction with $MoS_2$ intercalation.

**S5. Raman spectrum of multilayer $MoS_2$**

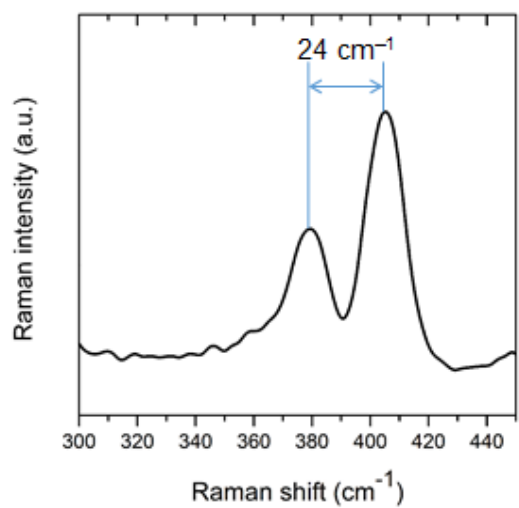

**Figure S5.** Raman spectra of multilayer MoS$_2$ before being transferred to the device. The peak position difference of E$^1_{2g}$ and A$_{1g}$ is ~24 cm$^{-1}$, indicating that the MoS$_2$ sheet is of ~4 layer thickness.[S6]

**S6. Richardson plot of short-circuiting photocurrent of the graphene/Si diode with MoS$_2$ intercalation**

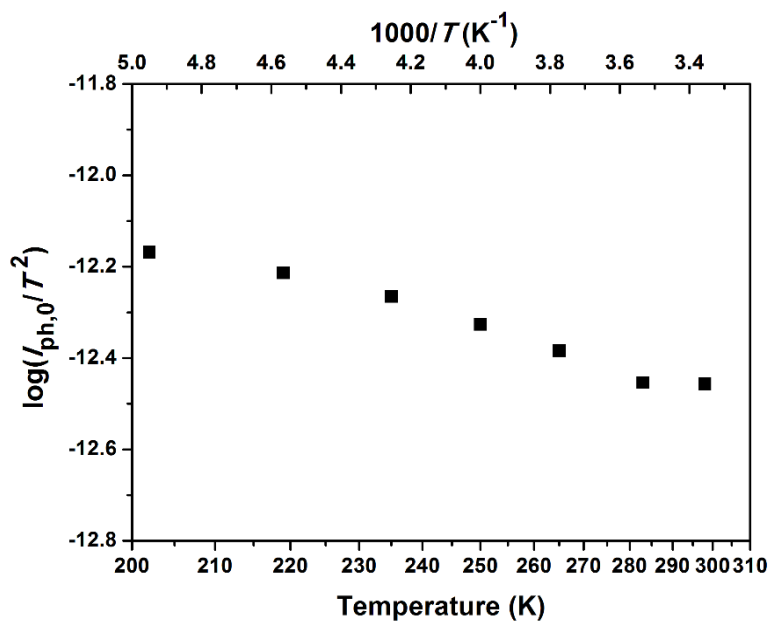

**Figure S6.** Richardson plot of $I_{ph,0}$ (log($I_{ph,0}/T^2$)-1/$T$ plot) in graphene/Si diode with MoS$_2$ intercalation.

## S7. Flexible hybrid graphene photoconductor

The fabrication process of the flexible device is very similar to that of rigid device. The major difference is the usage of Si thin film exfoliated from SOI rather than bulk Si. Firstly, SOI is put into HF solution (HF:H$_2$O=1:4) for etching SiO$_2$. Then the Si thin film (thickness ~200 nm) is transferred onto the polyimide substrate and patterned to rectangle by CF$_4$ etching. The following steps, e.g., transfer and patterning of MoS$_2$ film, photolithography and electrode deposition are the same with those of rigid device, as described in our main manuscript. **Figure S7** summarizes the schematic diagram and the photoresponse performance of the flexible device.

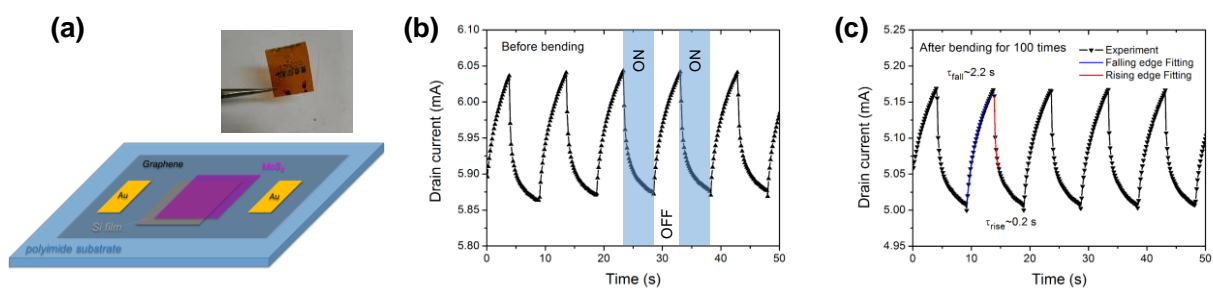

**Figure S7.** (a) Schematic diagram of the flexible photodetector on polyimide substrate using Si thin film exfoliated from silicon-on-insulator and optical image of the device. Photoresponse of the flexible device under 635 nm illumination (168 nW power, 5 V drain voltage) before (b) and after bending the device for 100 times (c). The red and blue curves are the exponential fitting for rising edge and falling edge, respectively.

## Supplementary references